\documentclass[lettersize,journal]{IEEEtran}
\usepackage{amsmath,amsfonts}
\usepackage{algorithmic}
\usepackage{algorithm}
\usepackage{array}
\usepackage[caption=false,font=normalsize,labelfont=sf,textfont=sf]{subfig}
\usepackage{textcomp}
\usepackage{stfloats}
\usepackage{url}
\usepackage{verbatim}
\usepackage{graphicx}
\usepackage{cite}
\usepackage{multirow}
\usepackage{multicol}
\usepackage{pgfplots}
\usepackage{booktabs}
\usepackage{amsmath}
\usepackage{tikz}
\usetikzlibrary{positioning}
\usepackage{hyperref} 
\newcommand{\rev}[1]{\textcolor{black}{#1}}
\newcommand{\etal}{\emph{et al.}}

\begin{document}

\title{Conversational Speech Recognition by Learning Audio-textual Cross-modal Contextual Representation}

\author{Kun Wei~\IEEEmembership{Student member,~IEEE}, Bei Li, Hang Lv, Quan Lu, Ning Jiang, Lei Xie,~\IEEEmembership{Senior member,~IEEE}
\thanks{Corresponding author: Lei Xie.}

\thanks{Kun Wei, Hang Lv and Lei Xie are with
the Audio, Speech and Language Processing Group, School of Computer Science, Northwestern Polytechnical
University, Xi’an 710072, China. Email: ethanwei@mail.nwpu.edu.cn (Kun Wei), hanglv@nwpu-aslp.org (Hang Lv), lxie@nwpu.edu.cn (Lei Xie).}

\thanks{Bei Li is with the School of Computer Science and Engineering, Northeastern University, Shenyang 110167, China. Email: libei\_neu@outlook.com.} 

\thanks{Quan Lu and Ning Jiang are with Mashang Consumer Finance Co., Ltd., Chongqing 401121, China. Email: quan.lu02@msxf.com (Quan Lu), ning.jiang02@msxf.com (Ning Jiang).}} 


\markboth{Journal of \LaTeX\ Class Files,~Vol.~14, No.~8, August~2021}%
{Shell \MakeLowercase{\textit{et al.}}: A Sample Article Using IEEEtran.cls for IEEE Journals}


\maketitle

\begin{abstract}

Automatic Speech Recognition (ASR) in conversational settings presents unique challenges, including extracting relevant contextual information from previous conversational turns. Due to irrelevant content, error propagation, and redundancy, existing methods struggle to extract longer and more effective contexts. To address this issue, we introduce a novel Conversational ASR system, extending the Conformer encoder-decoder model with cross-modal conversational representation. Our approach leverages a cross-modal extractor that combines pre-trained speech and text models through a specialized encoder and a modal-level mask input. This enables the extraction of richer historical speech context without explicit error propagation. We also incorporate conditional latent variational modules to learn conversational-level attributes such as role preference and topic coherence. By introducing both cross-modal and conversational representations into the decoder, our model retains context over longer sentences without information loss, achieving relative accuracy improvements of 8.8\% and 23\% on Mandarin conversation datasets HKUST and MagicData-RAMC, respectively, compared to the standard Conformer model.
\end{abstract}

\begin{IEEEkeywords}
Conversational ASR, Cross-modal Representation, Context, Conformer, Latent Variational.
\end{IEEEkeywords}

\section{Introduction}

\begin{figure}[t]
\centering
\includegraphics[scale=0.4]{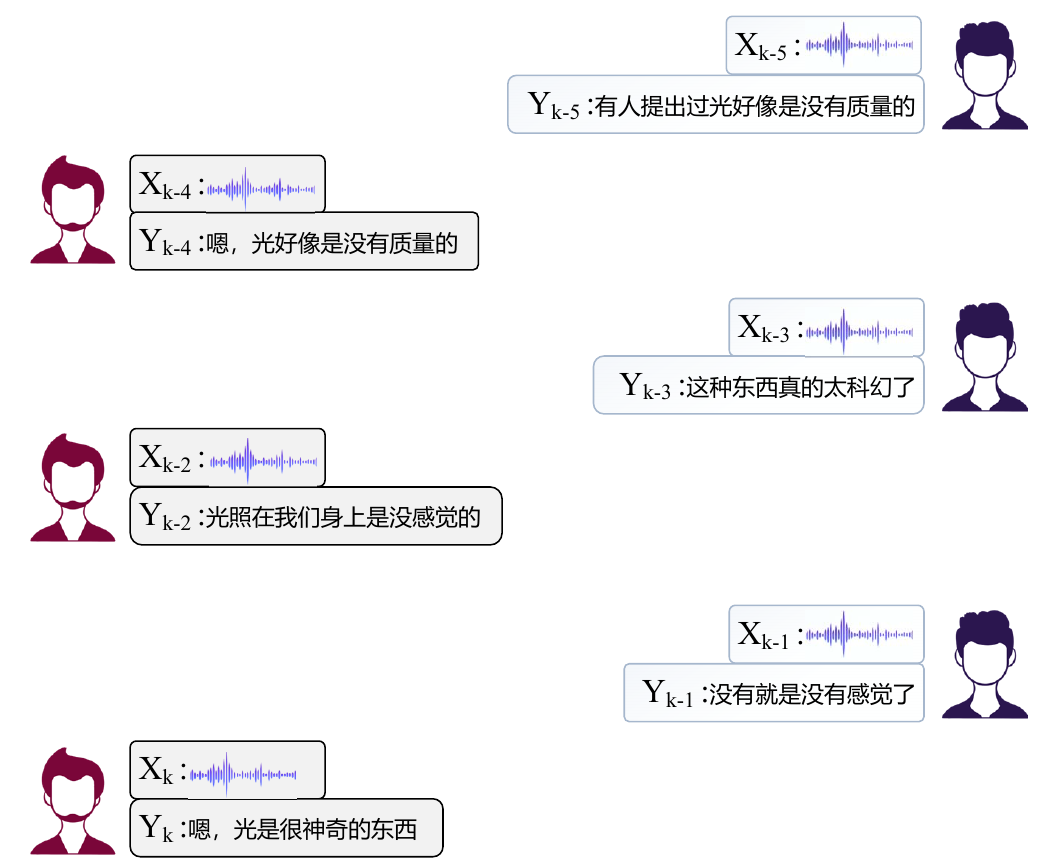}
\caption{An example of a conversation, where $X_k$ and $Y_k$ represent the speech and text of the current sentence $k$, respectively.}
\label{fig:speaker}
\end{figure}

\IEEEPARstart{A}{utomatic} Speech Recognition (ASR) has conventionally been designed for sentence-level transcription, leveraging paired sentence-level speech-text data for training purposes~\cite{hinton2012deep, li2022recent}. Nevertheless, the burgeoning demand for voice-activated interfaces in diverse applications such as meeting transcription and spoken dialog systems necessitates an ability to process extended, conversational speech as shown in Fig.~\ref{fig:speaker}. This form of speech introduces unique characteristics, including role-specific lexical preferences and context-dependent topical coherence~\cite{liang2021modeling, xiong2017toward}. Specifically, the above characteristics refer to the impact of conversational roles on the probability of certain words and phrases, and the influence of topic and discourse structure on the co-occurrence of semantically related words across adjacent sentences. Previous research indicates that incorporating contextual elements from prior utterances significantly augments conversational speech recognition performance~\cite{kim2019cross}.

Recent years have witnessed remarkable advances in end-to-end ASR architectures, including Connectionist Temporal Classification (CTC)~\cite{graves2012connectionist}, Recurrent Neural Network Transducer (RNN-T)~\cite{graves2012sequence}, and Attention-Based Encoder-Decoder (AED) models~\cite{vaswani2017attention, gulati2020conformer, kim2017joint}. These have shown substantial performance gains over traditional hybrid models~\cite{li2022recent}. However, effectively integrating extended contextual information into these models is a persistent challenge. Current solutions fall into three primary categories: 1) Text-based methods leverage language models to extract high-level textual features~\cite{mikolov2012context, mikolov2010recurrent}, sometimes employing auxiliary techniques like Variational Autoencoders (VAE)~\cite{wei2022conversational}. 2) Speech-based strategies establish a direct linkage between input speech and transcriptions at the sentence level~\cite{kim2018dialog, hori2020transformer, hori2021advanced}, or extracting speech context features using additional encoders~\cite{shon2023context, kojima2021large}. 3) Hybrid approaches incorporate both textual and speech-based features~\cite{wei2022conversational,gong2022longfnt,hou2022bring,cui2023towards,wei2022improving}. 

While existing methods attempt to incorporate historical context into current ASR tasks, they face inherent limitations in achieving optimal accuracy. Specifically, text-based approaches can easily capture longer context but also introduce a mismatch between training and inference stages, causing errors in historical sentence recognition to propagate into the inference of the current sentence. Meanwhile, speech-based approaches are more soft and realistic but obviously introduce redundant information, thereby diverting the model's focus from relevant features~\cite{wei2022conversational}. Although hybrid methods attempt to combine the advantages of both context text and speech, they inevitably integrate the drawbacks of the two~\cite{wei2022improving}. Due to the current hybrid approaches primarily introducing speech and acoustic information separately in different modules, it also leads to the simultaneous introduction of error transmission problems and overly abundant additional features in speech when utilizing local information and longer text information in conversations. Despite attempts to amalgamate these approaches, existing methods still fall short of effectively leveraging longer contextual information. Consequently, there is an unmet need for a technique that improves the extraction of longer contextual information and mitigates error propagation and attention dilution. Or rather, we need to extract longer and more effective context at the same time from the conversations.

To address this issue, we introduce a \rev{novel} ASR model based on an attention-based Conformer encoder-decoder~\cite{gulati2020conformer}, augmented with a Conditional Variational Autoencoder (CVAE) for cross-modal representation. We employ cross-modal features to extract conversation-level representations in longer contexts, implicitly utilizing contextual information, thereby avoiding the error propagation problems brought by overly long text information. By combining local cross-modal and long-context conversational representations, we aim to use longer and more accurate conversational contexts to improve speech recognition performance. Specifically, this architecture leverages pre-trained models, such as data2vec~\cite{baevski2022data2vec} and HuBERT~\cite{hsu2021hubert} for speech, and RoBERTa-wwm-ext~\cite{liu2019roberta} for text, to extract cross-modal representations conducive to downstream tasks. The model is trained to capture local context dependencies through L1 loss and CTC loss, while role-specific and topic-specific variational modules are employed to refine the conversational context. In the process of recognizing conversation speech, the cross-modal representation of the current sentence and CVAE conversational representations are concatenated and sent into the decoder, introducing both local and long context into the speech recognition framework. Our framework substantially improves ASR performance, attaining up to 23\%  improvements on the test datasets.

Our contributions are threefold:
\begin{itemize}
    \item We present a novel ASR framework that integrates cross-modal representation and a CVAE module, enhancing the model's ability to contextualize conversational speech.
    \item Our method demonstrates a significant decrease in ASR error rate, achieving up to 8.8\% and 23\% character error rate reduction on the HKUST and MagicData-RAMC datasets, respectively.
    \item We investigate the influence of different pre-trained models and input lengths on the performance, establishing an optimized CVAE input configuration through empirical analysis.
\end{itemize}

\rev{Building on previous work~\cite{wei2022conversational}, we integrate the CVAE module's ability to extract extended contexts with the capacity of cross-modal extractors~\cite{kun2022Leveraging} to obtain more precise contextual representations. In contrast to our prior research, we extend the CVAE to new modalities, investigate various fusion strategies of the Decoder for context information, and incorporate additional context information to enhance the model's ability to utilize both global and local contexts. Moreover, through a series of experiments, we examine the CVAE model's performance under new modalities and its influence on ASR recognition capabilities. The expansion of the input mode for conversational speech bolsters the framework's ability to extract and utilize conversational representations.}

\begin{figure*}[htbp]
\centering
\tikzstyle{netnode}=[rounded corners=5pt, fill=cyan!80!black!20, inner sep=0pt, minimum width=5.5em, minimum height=2.2em]
\tikzstyle{encodernode}=[rounded corners=5pt, fill=yellow!40, inner sep=0pt, minimum width=12em, minimum height=2.2em]
\tikzstyle{wnode}=[rounded corners=5pt, inner sep=5pt, minimum width=2em, minimum height=1.8em, draw, fill=black!5!white]
\tikzstyle{cnode}=[circle,minimum size=2em, inner sep=0pt, draw, fill=black!5!white]
\tikzstyle{archnode}=[encodernode, minimum width=11em]
\vspace{-0.3cm}
\scalebox{0.9}{
\begin{tikzpicture}

\node[anchor=center,netnode](Postnet_r_ori) at (0,0) {};
\node[anchor=west,draw,dashed,rounded corners=10pt, minimum width=4.65in, minimum height=3.6in, fill=black!20!white,opacity=0.1](bg) at ([xshift=-0.8em,yshift=-0.46in]Postnet_r_ori.west) {};

\node[anchor=west,draw,dashed,rounded corners=10pt, minimum width=1.15in, minimum height=3.6in, fill=black!20!white,opacity=0.1](bg2) at ([xshift=33.2em,yshift=-0.46in]Postnet_r_ori.west) {};

\node[anchor=north west, align=left](Postnet_r) at ([xshift=0.5em,yshift=-0.5em]bg2.north west) {\textbf{CRM}\\\textbf{Module}};

\node[anchor=north west, align=left](Postnet_r) at ([xshift=0.5em,yshift=-0.5em]bg.north west) {\textbf{CVAE}\\\textbf{Module}};

\node[anchor=north east, align=left](Postnet_r) at ([xshift=-19.0em,yshift=-23em]bg.north east) {\text{LVM$_{\text{role}}$=Prenet$_{\text{role}}$+Postnet$_{\text{role}}$}};

\node[anchor=north east, align=left](Postnet_r) at ([xshift=-17.4em,yshift=-24em]bg.north east) {\text{LVM$_{\text{topical}}$=Prenet$_{\text{topical}}$+Postnet$_{\text{topical}}$}};

\node[anchor=center,netnode](Postnet_r) at (0,0) {Postnet$_{\text{role}}$};
\node[anchor=west,netnode](Postnet_t) at ([xshift=1em]Postnet_r.east) {Postnet$_{\text{topical}}$};
\node[anchor=west,netnode](Prenet_r) at ([xshift=1.5em]Postnet_t.east) {Prenet$_{\text{role}}$};
\node[anchor=west,netnode](Prenet_t) at ([xshift=4.6em]Prenet_r.east) {Prenet$_{\text{topical}}$};

\node[anchor=north,encodernode](LVM1) at ([xshift=0.5em, yshift=-3em]Postnet_r.south east) {Text Encoder};
\node[anchor=north west,encodernode, text width=17em, minimum height=3.6em,align=center](CME) at ([xshift=0em,yshift=-3em]Prenet_r.south west) 
  {Feature Module \\ (Cross-modal Extractor/Text Encoder)};

{\footnotesize

\node[anchor=center](strole) at ([xshift=-4em, yshift=0.6em]CME.north) {$S_{role}$/$S^{text}_{role}$};
\node[anchor=center](sttopical) at ([xshift=-1em, yshift=0.6em]CME.north east) {$S_{topical}$/$S^{text}_{topical}$};

}

\node[anchor=north,wnode](TargetText) at ([yshift=-2em]LVM1.south) {$Y_{k}$};

\node[anchor=north west,wnode](Speech_r) at ([xshift=0.2em, yshift=-1.5em]CME.south west) {X$_{role}$/Y$_{role}$};
\node[anchor=west,wnode](Speech_t) at ([xshift=3.3em]Speech_r.east) {X$_{topical}$/Y$_{topical}$};



{\footnotesize
\node[anchor=west,cnode] (vcon) at  ([xshift=9em]Prenet_t.east) {V$_{con}$};
\node[anchor=south,cnode] (vpr) at ([xshift=0pt, yshift=1em]Postnet_r.north) {V$^{p}_r$};
\node[anchor=south,cnode] (vpt) at ([xshift=0pt, yshift=1em]Postnet_t.north) {V$^{p}_t$};
\node[anchor=south,cnode] (vr) at ([xshift=0pt, yshift=1em]Prenet_r.north) {V$_{r}$};
\node[anchor=south,cnode] (vt) at ([xshift=0pt, yshift=1em]Prenet_t.north) {V$_{t}$};
}

\node[anchor=north,archnode, minimum width=7em, text width=6em,  align=center, fill=yellow!40](CE2) at ([xshift=0pt, yshift=-3.2em]vcon.south) {Cross-modal \\ Extractor};
\node[anchor=center](stc) at ([xshift=2em, yshift=0.6em]CE2.north) {$S_{context}$};
\node[anchor=north,wnode](Speech_c) at ([xshift=0em, yshift=-2.0em]CE2.south) {X$_{context}$};

\node[anchor=west,archnode, fill=cyan!80!black!20](CD) at ([xshift=3.5em, yshift=0pt]vcon.east) {Conditional Decoder};
\node[anchor=north,archnode, fill=yellow!40](CE) at ([xshift=0pt, yshift=-2em]CD.south) {Conformer Encoder};
\node[anchor=north,archnode, fill=green!60!black!20](SPM2) at ([xshift=0pt, yshift=-2.0em]CE.south) {Speech Pretrained model};
\node[anchor=south,archnode, fill=black!5!white](Softmax) at ([xshift=0pt, yshift=1.5em]CD.north) {Softmax};

\node[anchor=south,wnode] (yk) at ([xshift=0pt, yshift=1em]Softmax.north) {$Y_{k}$};

\node[anchor=north,wnode](Speech_k) at ([xshift=0em, yshift=-2.0em]SPM2.south) {X$_{k}$};

\node[anchor=south,align=center,wnode](KLnode) at ([xshift=0.25em, yshift=4em]Postnet_t.north east) {$KL(q_{\phi}(\textbf{V}^{p}_\text{role})||p_{\theta}(\textbf{V}_\text{role}))$ \\
 $KL(q_{\phi}(\textbf{V}^{p}_\text{topic})||p_{\theta}(\textbf{V}_\text{topical}))$};

{\scriptsize
\node[anchor=north west,align=left](stringeqx) at ([xshift=-2em, yshift=-0.3em]Speech_r.south west) {$\ldots,X_{k-4},X_{k-2}=X_{role}$\\$\ldots,Y_{k-4},Y_{k-2}=Y_{role}$};
\node[anchor=north,align=center](stringeqxc) at ([xshift=0.0em, yshift=-2.0em]Speech_c.south) {$X_{k-1},X_{k}=X_{context}$};
\node[anchor=north west,align=left](stringeqy) at ([xshift=12.4em, yshift=-0.3em]Speech_r.south west) {$\ldots,X_{k-2},X_{k-1}=X_{topical}$\\$\ldots,Y_{k-2},Y_{k-1}=Y_{topical}$};
}

{
\draw[->, rounded corners] (TargetText.north) -- (LVM1.south);

\draw[->, rounded corners] ([xshift=0em, yshift=-3em]Postnet_r.south) -- (Postnet_r.south);
\draw[->, rounded corners] ([xshift=0em, yshift=-3em]Postnet_t.south) -- (Postnet_t.south);

\draw[->, rounded corners] ([xshift=0pt, yshift=0pt]Postnet_r.north) -- (vpr.south);
\draw[->, rounded corners] ([xshift=0pt, yshift=0pt]Postnet_t.north) -- (vpt.south);
}

{
\draw[->, rounded corners] (Speech_r.north) -- ([xshift=0em, yshift=1.5em]Speech_r.north);
\draw[->, rounded corners, draw=orange] (CE2.north) -- ([xshift=0em, yshift=0.0em]vcon.south);
\draw[->, rounded corners] (Speech_t.north) -- ([xshift=0em, yshift=1.5em]Speech_t.north);
\draw[->, rounded corners] (Speech_c.north) -- ([xshift=0em, yshift=0.0em]CE2.south);

\draw[->, rounded corners] ([xshift=0em, yshift=5.1em]Speech_r.north) -- ([xshift=0em, yshift=8.1em]Speech_r.north);
\draw[->, rounded corners] ([xshift=0em, yshift=5.1em]Speech_t.north) -- ([xshift=0em, yshift=8.1em]Speech_t.north);

\draw[->, rounded corners] ([xshift=0pt, yshift=0pt]Prenet_r.north) -- ([xshift=0pt, yshift=0pt]vr.south);
\draw[->, rounded corners] ([xshift=0pt, yshift=0pt]Prenet_t.north) -- ([xshift=0pt, yshift=0pt]vt.south);
}

{

}

{

\draw[->, rounded corners=2pt, draw=orange] ([xshift=0em, yshift=0em]vr.east) -- ([xshift=1em, yshift=0em]vr.east) -- ([xshift=1em, yshift=-1.3em]vr.east) -- ([xshift=-0.1em, yshift=0.3em]Prenet_t.north)  .. controls + (135:0.3em) and +(45:0.3em) ..  ([xshift=0.1em, yshift=0.3em]Prenet_t.north) -- ([xshift=1em, yshift=0.3em]Prenet_t.north east) -- ([xshift=1.0em, yshift=0em]Prenet_t.east) -- ([xshift=0em, yshift=0em]vcon.west);
\draw[->, rounded corners=2pt, draw=orange] ([xshift=0em, yshift=0em]vt.east) -- ([xshift=3.75em, yshift=0em]vt.center) -- ([xshift=1.0em, yshift=0em]Prenet_t.east) -- ([xshift=0em, yshift=0em]vcon.west);
\draw[->, rounded corners=2pt, draw=orange] ([xshift=0em, yshift=0em]vcon.east) -- ([xshift=0em, yshift=0em]CD.west);
}

{
\draw[->] ([xshift=0em, yshift=0em]vpr.north) ..controls + (north:1.7em) and + (south:1.5em).. ([xshift=0em, yshift=0em]KLnode.south);
\draw[->] ([xshift=0em, yshift=0em]vpt.north) ..controls + (north:1.4em) and + (south:1.6em).. ([xshift=0em, yshift=0em]KLnode.south);
\draw[->] ([xshift=0em, yshift=0em]vr.north) ..controls + (north:1.4em) and + (south:1.6em).. ([xshift=0em, yshift=0em]KLnode.south);
\draw[->] ([xshift=0em, yshift=0em]vt.north) ..controls + (north:1.7em) and + (south:1.5em).. ([xshift=0em, yshift=0em]KLnode.south);
}

{
\draw[->, rounded corners] ([xshift=0pt, yshift=0pt]Softmax.north) -- ([xshift=0pt, yshift=0pt]yk.south);
\draw[->, rounded corners] ([xshift=0pt, yshift=0pt]Speech_k.north) -- ([xshift=0pt, yshift=0pt]SPM2.south);
\draw[->, rounded corners] ([xshift=0pt, yshift=0pt]SPM2.north) -- ([xshift=0pt, yshift=0pt]CE.south);
\draw[->, rounded corners] ([xshift=0pt, yshift=0pt]CE.north) -- ([xshift=0pt, yshift=0pt]CD.south);
\draw[->, rounded corners] ([xshift=0pt, yshift=0pt]CD.north) -- ([xshift=0pt, yshift=0pt]Softmax.south);
}

\end{tikzpicture}
}
\caption{The framework of the CVAE-based conversational ASR. In this figure, X represents the speech input. 
The CVAE module comprises a target text encoder and two Latent Variational Modules (LVM). During the training process, the output from the Postnet is sent to the decoder. Conversely, during the decoding process, the output of the Prenet is utilized. For training purposes, $\textbf{V}^{p}_\text{role}, \textbf{V}^{p}_\text{topical}$ are employed, while $\textbf{V}_\text{role}, \textbf{V}_\text{topical}$ are used for decoding. In this figure, $\textbf{V}_{con}$ represents $\textbf{V}_{context}$. The two text encoders in the CVAE module share model parameters. Moreover, the cross-modal extractor in both the CVAE Module and the CRM Module also share model parameters.
}
\vspace{-0.3cm}
\label{fig:frame}
\end{figure*}
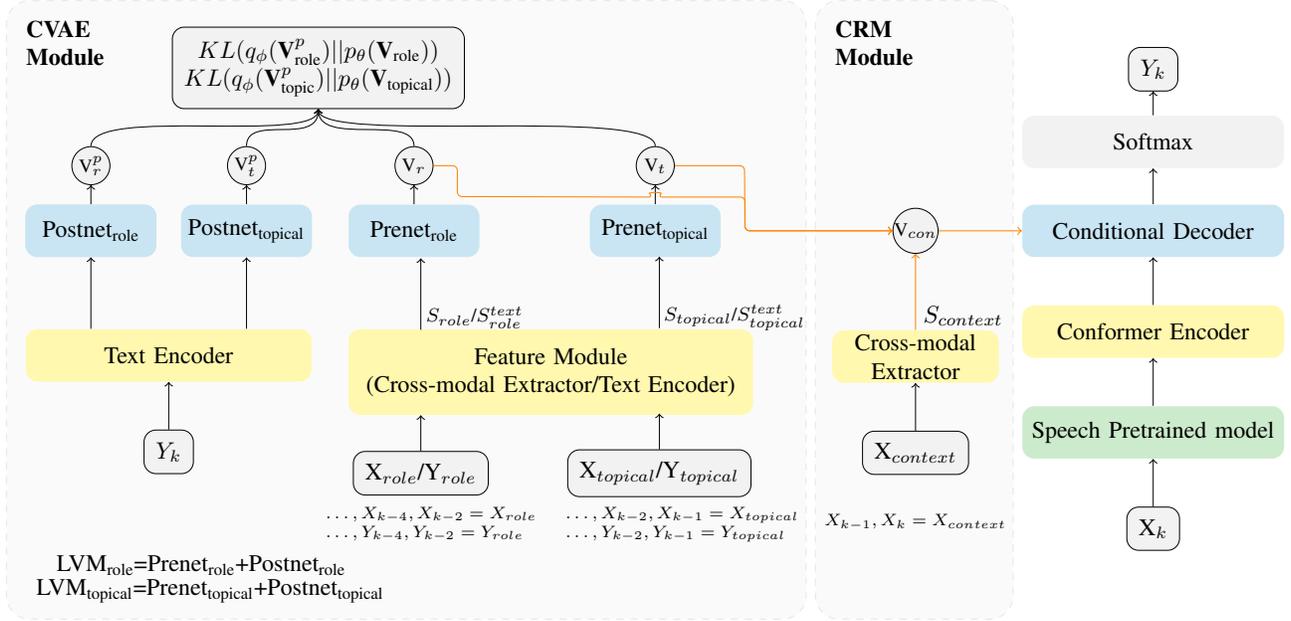

\section{Cross-modal CVAE Based Conversational Speech Recognition}
Our model consists of a Conformer encoder, a cross-modal extractor, a conversational representation extractor (CRM), and a conditional decoder. As shown in Fig.~\ref{fig:frame}, the features extracted from the speech pre-trained model will be simultaneously fed into the cross-modal extractor and the Conformer encoder when training. The cross-modal representation of the context is then fed into the CVAE module to generate two conversational representations, namely topical coherence representation and role preference representation. These two conversational representations are then integrated into the decoding process of speech recognition through the fusion modules, ultimately helping the speech recognition model obtain conversational context information. In other words, the conversational decoder gets the final recognition result by fusing the output representation and conversational representation of the Conformer encoder. In this section, we will introduce the composition of each module in detail.

\subsection{Input Representation}
When we aim to recognize speech utterance $\textbf{X}_k$, we define $\textbf{X}_{topical}$ as the speech of several consecutive preceding sentences and $\textbf{Y}_{topical}$ as the text of several consecutive preceding sentences to obtain the topical coherence information of the conversation. Here, $\textbf{X}_{topical} = (...,\textbf{X}_{k-4},\textbf{X}_{k-3},\textbf{X}_{k-2},\textbf{X}_{k-1})$, $\textbf{Y}_{topical} = (...,\textbf{Y}_{k-4},\textbf{Y}_{k-3},\textbf{Y}_{k-2},\textbf{Y}_{k-1})$. For example, when the local coherence length is defined as 3, the topical formula can be expressed as: $\textbf{X}_{topical} = (\textbf{X}_{k-3},\textbf{X}_{k-2},\textbf{X}_{k-1})$ and $\textbf{Y}_{topical} = (\textbf{Y}_{k-3},\textbf{Y}_{k-2},\textbf{Y}_{k-1})$. Simultaneously, we establish a representation of the role information, where $\textbf{X}_{role}$ represents the speech of the current speaker's previous n sentences, and $\textbf{Y}_{role}$ denotes the text of the current speaker's previous n sentences. When the role information length is defined as 3, the role formula can be expressed as: $\textbf{X}_{role} = (\textbf{X}_{k-6},\textbf{X}_{k-4},\textbf{X}_{k-2})$ and $\textbf{Y}_{role} = (\textbf{Y}_{k-6},\textbf{Y}_{k-4},\textbf{Y}_{k-2})$. Furthermore, the cross-modal extractor extracts the cross-modal representations from the current and preceding speech utterances ($\textbf{X}_{k-1},\textbf{X}_{k}$), denoted as $\textbf{X}_{context}$.



\subsection{Conformer Encoder}

In our framework, the Conformer encoder accepts the features generated by the speech pre-training model and outputs the intermediate representation $\textbf{z}$ of the current speech to be recognized. As one of the most advanced end-to-end speech recognition architectures available, the Conformer encoder is constructed using a series of Conformer blocks, each containing a convolution module, a multi-headed self-attention module, and two feed-forward modules. The self-attention module captures global contextual information from the input speech, while the convolution layer focuses on extracting local correlations.

The Conformer encoder consists of a convolutional feature extractor and several interconnected Conformer blocks. Given an input speech feature sequence $\widetilde{\textbf{X}}_{i}$ (extracted from $\textbf{X}_{i}$), it is first passed through the convolutional down-sampling module, which yields a dimensionality-reduced feature. Subsequently, the features serve as the input for the concatenated Conformer blocks, resulting in the encoder output $\textbf{z}$.

For a given layer with input $\widetilde{\textbf{X}_i}$, the input sequentially passes through a feed-forward (FFN) module, a multi-head self-attention (MHSA) module, a convolution (CONV) module, and another feed-forward module to produce the output of the block.

The FFN module comprises two linear layers and a nonlinear activation layer. Like the Transformer model~\cite{dong2018speech}, the module includes residual connections and layer normalization. In this model, the nonlinear activation function utilized is the Swish activation~\cite{ramachandran2017searching}. The MHSA module integrates the relative sinusoidal positional encoding scheme~\cite{dai2019transformer}. The CONV module begins with a gating mechanism~\cite{dauphin2017language}, followed by a one-dimensional convolution layer and batch normalization.

To further elaborate, the computational process of a Conformer block consists of the following components: 
\begin{align} 
\label{Conformer}
\hat{\textbf{X}}_{i} &= \widetilde{\textbf{X}} + \frac{1}{2}{\rm FFN}(\widetilde{\textbf{X}}_{i}), \\
\overline{\textbf{X}}_{i} &= {\rm MHSA}(\hat{\textbf{X}}_{i}) + \hat{\textbf{X}}_{i},\\
\textbf{X}^{'}_{i} &= {\rm CONV}(\overline{\textbf{X}}_{i}),\\
\textbf{C}_{i} &= {\rm Layernorm}(\frac{1}{2}{\rm FFN}(\textbf{X}^{'}_{i}) + \textbf{X}^{'}_{i}).
\end{align}

The calculated output, $\textbf{C}_{i}$, is the subsequent Conformer block layer input. In our framework, the encoder accepts the features output by the speech pre-trained model and outputs the intermediate representation $\textbf{z}$ of the current speech to be recognized.

\begin{figure}[]
\centering
\includegraphics[scale=0.5]{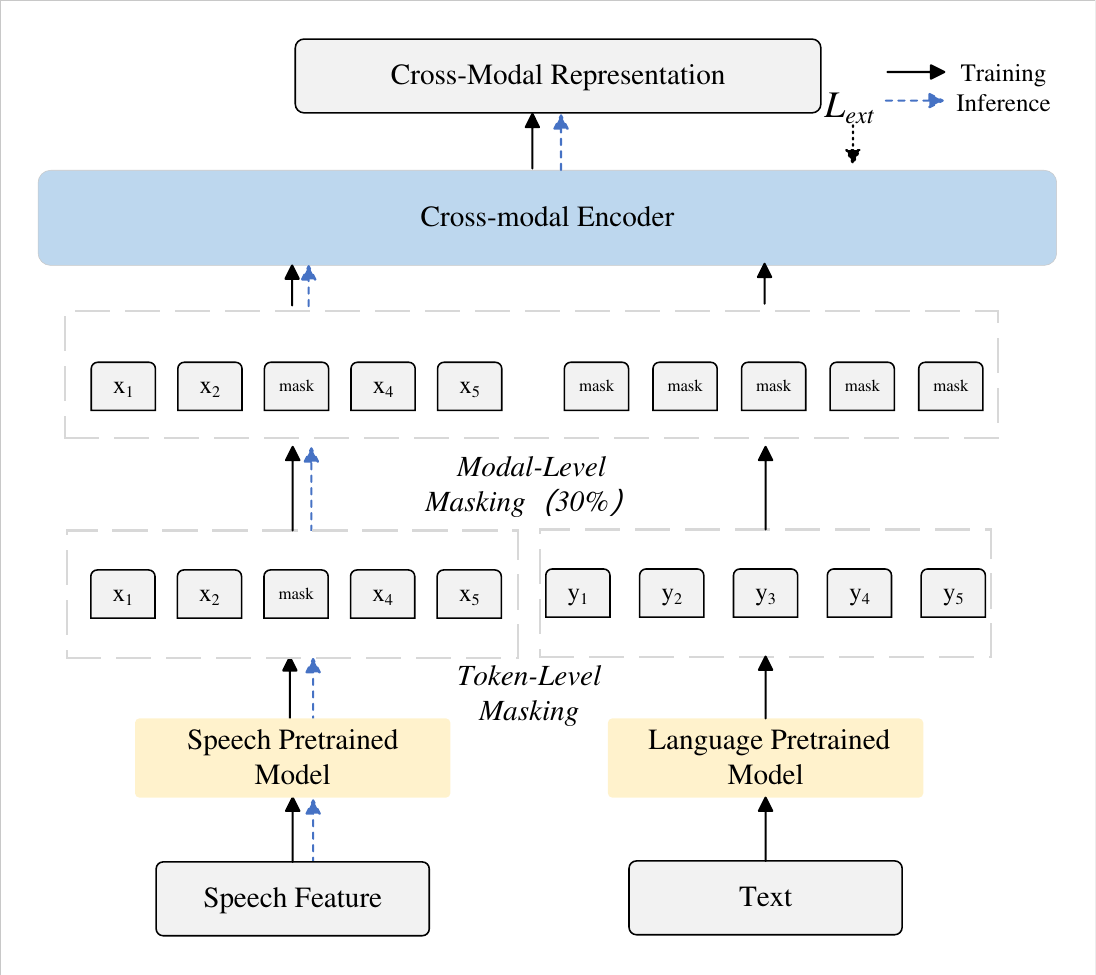}
\caption{Framework of the cross-modal extractor. Either the speech or text modality will be randomly masked. \textit{mask} represents the masked token. The black and blue lines in the model represent the training and inference paths, respectively.}
\vspace{-0.3cm}
\label{fig:cross-modal}
\end{figure}

\subsection{Cross-modal Extractor}
We use a cross-modal extractor to extract linguistic information from speech. The cross-modal extractor employs only the speech features $\widetilde{\textbf{X}}_{context}, \widetilde{\textbf{X}}_{role}$, and $\widetilde{\textbf{X}}_{topical}$ during conversational speech recognition. We use a pre-trained speech model to extract essential information from the speech input while concurrently filtering out redundant information. At the same time, we can also use the linguistic information to help the conversational speech recognition model obtain more accurate context representations of speech features $\textbf{S}_{context}, \textbf{S}_{role}$ and $\textbf{S}_{topical}$. The details of the cross-modal extractor will be introduced in Section III.
\subsection{CVAE Based Conversational ASR}
While utilizing local context, we use cross-modal representations $\textbf{S}_{role}$ and $\textbf{S}_{topical}$ to extract longer conversational representations $\textbf{V}_{role}, \textbf{V}_{topical}$. By only using cross-modal representations generated from historical speech, we avoid explicit error propagation and introduce more helpful context into the speech recognition process. \rev{The CVAE module comprises a target text encoder and two Latent Variational Modules (LVM), each LVM is composed of a Prenet and a Postnet.} The process of extracting cross-modal conversation representations using the CVAE will be detailed in Section IV.

\subsection{Conditional Decoder}
\begin{figure}[htbp]
\centering
\includegraphics[scale=0.5]{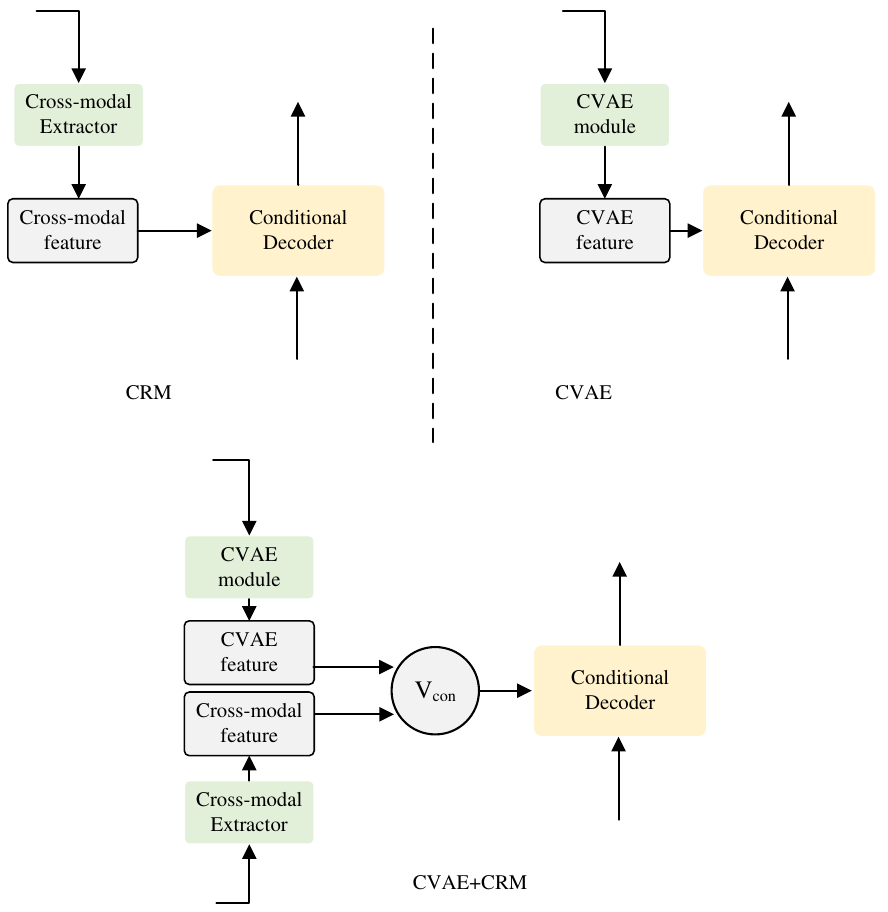}
\caption{Different decoding strategies: $\textbf{V}_{context}=\textbf{S}_{context}$ in CRM, $\textbf{V}_{context}=(\textbf{V}_{role},\textbf{V}_{topical})$ in CVAE, and $\textbf{V}_{context}=(\textbf{V}_{role},\textbf{V}_{topical}, \textbf{S}_{context})$ in CRM+CVAE.}
\label{fig:crm}
\end{figure}

We explore two strategies to integrate conversation representations into the ASR model: adding an additional attention layer to the decoder (Attention Condition) and splicing the output vector directly (Linear Condition). As shown in Fig.~\ref{fig:crm}, suppose we obtain the output $\textbf{V}_{role}, \textbf{V}_{topical}$ from the LVMs, the input of conditional decoder can be $\textbf{V}_{context}=(\textbf{V}_{role}, \textbf{V}_{topical})$ or $\textbf{V}_{context}=(\textbf{V}_{role}, \textbf{V}_{topical}, \textbf{S}_{context})$. We describe the two fusion strategies in detail below. 


\subsubsection{Attention Condition}
In the traditional framework, the decoder closely resembles the Transformer model~\cite{vaswani2017attention}, with a notable distinction in the multi-headed attention layer. 
Given the target text feature $\textbf{q}_l$, the computational process for the $l$-th block in the decoder proceeds as follows:
\begin{align}
    \hat{\textbf{q}}_{l} &= {\rm MHSA}(\textbf{q}_{l}) + \textbf{q}_{l}, \\
    \textbf{q}_{l+1} &= {\rm MHA}(\hat{\textbf{q}}_{l},\textbf{z}) + \hat{\textbf{q}}_{l},
\end{align}
where $\bf{z}$ denotes the output feature from the final layer of the Conformer encoder, while MHA represents the multi-head attention module.

 We first attempt to add an attention layer parallel to the encoder output at each decoder layer. Specifically, the structure of each decoder layer is as follows:
\begin{align}
    \hat{\textbf{q}}_{l} &= {\rm MHSA}(\textbf{q}_{l}) + \textbf{q}_{l}, \\
    \textbf{p}_{l} &= {\rm MHA}(\hat{\textbf{q}}_{l},\textbf{z}) + \hat{\textbf{q}}_{l}, \\
    \textbf{q}_{l+1} &= {\rm MHA}(\textbf{p}_{l},\textbf{V}_{context}) + \textbf{p}_{l}.
\end{align}

Finally, the output vector of the decoder blocks will be sent to the Softmax layer to calculate each word's occurrence probability.

\subsubsection{Linear Condition}
While using an attention mechanism can fully integrate context information into the decoding process, as the context length gradually increases, especially when it increases to several times the speech to be recognized, the attention mechanism will inevitably have its weights dispersed, leading to a deterioration in the final ASR recognition result under the same training step. Therefore, we explore another strategy to integrate conversation characteristics. In this approach, we only fuse the context information of the conversation at the output position of the decoder:
\begin{equation}
    \textbf{g}_{t} = \rm{Tanh}(\textbf{W}_\text{trans}(\textbf{V}_{context}, \textbf{q}_\text{L})+\textbf{b}_\text{trans}),
\end{equation}
where $\textbf{q}_\text{L}$ is the decoder state of the L layer, $\textbf{W}_\text{trans}$ and $\textbf{b}_\text{trans}$ are the weights and offsets of the linear layer, respectively.
$\textbf{g}_{t}$ will be sent to the Softmax layer for classification, and finally, the recognition probability of each word will be obtained.

\subsection{Training Objectives}

Following our previous work~\cite{wei2022conversational}, we first train a sentence-level speech recognition model based on the input of a single sentence. The training goal is to minimize the distance between the model output and the real transcript. Specifically, we use the cross-entropy loss as the objective function, which is defined as follows:
\begin{equation}
    \mathcal{L}_\text{CE}(\theta_{asr}; \textbf{X}, \textbf{Y}) = -\sum_{t=1}^n \log{p_{\theta_{asr}}(y_{t}|\textbf{X}, y_{1:t-1})}.
\end{equation}

Once the sentence-level model has learned enough information from individual sentences, we introduce role preference and topical coherence in the conversation to enhance its ability to recognize speech in a conversational setting. The training goal of the model at this stage is to jointly optimize the sentence-level ASR model and the LVMs using a multi-task learning framework:
\begin{equation}
\begin{aligned}
& \mathcal{L}_\text{final}(\theta, \phi; \textbf{V}_\text{role}, \textbf{V}_\text{topical}, \textbf{X}, \textbf{Y})  = \\                                       &+KL(q_{\phi}(\textbf{V}^{p}_\text{role}|\textbf{S}_\text{role},\textbf{Y}_{k})||p_{\theta}(\textbf{V}_\text{role}|\textbf{S}_\text{role})) \\
&+KL(q_{\phi}(\textbf{V}^{p}_\text{topic}|\textbf{S}_\text{topical},\textbf{Y}_{k})||p_{\theta}(\textbf{V}_\text{topical}|\textbf{S}_\text{topical}))\\
&-\mathcal{E}[\log p_{{\theta}asr}(y_{t}|\textbf{X}, y_{t-1}, \textbf{V}_\text{role}, \textbf{V}_\text{topical})].
\end{aligned}
\end{equation}  

When training the ASR model, the parameters of the speech pre-trained model and the cross-modal extractor will be frozen.
\section{The Cross-modal Extractor}
In addition to leveraging the Conformer encoder for speech feature extraction, we employ a pre-trained speech model as an input feature extractor for our cross-modal component. Our audio-textual cross-modal extractor facilitates the extraction of semantically aligned speech features.

As depicted in Fig.~\ref{fig:cross-modal}, this extractor comprises a pre-trained language model, a pre-trained speech model and a specialized cross-modal encoder. During training, paired speech features and their corresponding transcripts serve as input. Textual inputs, denoted as $\textbf{Y}_{topical}, \textbf{Y}_{role}$, are processed through the pre-trained language model to yield high-dimensional textual features. These features are combined with speech features in the cross-modal encoder, resulting in a representation enriched with both speech and semantic context. Notably, both the speech and language models are pre-trained using unsupervised data, enhancing their ability to capture richer contextual information while filtering out irrelevant noise. Upon obtaining the pre-trained features for speech and text, these are concatenated and fed into the cross-modal encoder to generate comprehensive cross-modal representations. To further refine this representation, we employ masking techniques during training: portions of the text and speech features are randomly masked, and the model is trained to predict these masked sections from the surrounding context. Moreover, we extend the masking strategy to the entire text or speech feature, enabling the model to learn the inter-modal correspondences better.

To mitigate the risk of explicit error propagation arising from contextual text inputs, our cross-modal extractor is designed to rely solely on speech features as input to the cross-modal encoder for generating cross-modal representations. This approach builds upon and enhances our previous work~\cite{wei2022conversational}.
Contrary to other cross-modal pre-training methodologies~\cite{wang2023understanding, bapna2021slam, zhang2022speechlm, zhou2022mmspeech}, our model exclusively utilizes data from downstream tasks during its training phase. This specificity reduces the likelihood of error and limits the additional parameter overhead to merely the size of the cross-modal encoder, significantly reducing the computational cost for its application in downstream tasks.
Subsequent subsections will delve into the specifics of the speech and language pre-trained models, the architecture of the cross-modal encoder, and the training objectives for the cross-modal extractor.

\subsection{Speech Pretrained Model}
Recently, the rapid advancement of unsupervised pre-training technology has led to the emergence of numerous novel speech pre-trained models, including wav2vec2.0~\cite{baevski2020wav2vec}, HuBERT, and data2vec. In comparison to wav2vec2.0, which was employed in~\cite{wei2022conversational}, HuBERT utilizes k-means clustering for modeling speech pre-trained models, while data2vec leverages a student model to predict masked speech input embeddings and learn context dependencies within sentences. Both models have demonstrated superior performance to wav2vec2.0 in downstream speech tasks~\cite{yang2021superb}.


Our work uses a speech pre-trained model augmented with a linear layer for speech feature extraction. We conducted experiments employing both HuBERT and data2vec, which were trained on the WenetSpeech dataset~\cite{zhang2021wenetspeech}. This dataset encompasses 10,000 hours of unsupervised Chinese speech data collected from the Internet. Consistent with the goals of our previous work~\cite{kun2022Leveraging}, we incorporate a linear layer to ensure dimensional compatibility between the output features of the speech and language pre-trained models. 

\subsection{Language Pretrained Model}

In the text encoding component of our architecture, we employ the RoBERTa-wwm-ext model~\cite{liu2019roberta, cui-etal-2021-pretrain}, a Chinese language pre-trained model that has been publicly released. This model incorporates a Whole Word Masking (wwm) strategy tailored for Chinese BERT and deviates from the traditional BERT model by eliminating the Next Sentence Prediction (NSP) task~\cite{devlin2019bert}. Further refinements in its training procedures have enabled RoBERTa-wwm-ext to excel in a diverse range of downstream tasks in Chinese natural language processing. The model has been trained on an extensive corpus of 5.4 billion tokens, encompassing a variety of domains such as news, encyclopedias, and question-answering platforms. Analogous to the design of the speech encoder, we append a linear layer to the output of the RoBERTa-wwm-ext model, mirroring the approach taken with our speech encoder.

\subsection{Cross-Modal Encoder}
The cross-modal encoder (CME) is designed to learn the correspondence between speech features and text features. We construct the cross-modal encoder using a three-layer Transformer block configuration~\cite{vaswani2017attention}. After acquiring the speech features $\widetilde{\textbf{X}}$ and the text features $\widetilde{\textbf{Y}}$, we concatenate the two features and feed them into the cross-modal encoder to obtain the final cross-modal context representation $\textbf{S}$:

\begin{equation}
\textbf{S} = \rm{CME}(\widetilde{\textbf{X}};\widetilde{\textbf{Y}}),
\end{equation}

\noindent where CME represents the cross-modal encoder and $(\cdot;\cdot)$ denotes the concatenation operation. To enhance the mutual learning capability between the modal features, the input text and speech features will be permuted in sequence randomly:

\begin{equation}
\textbf{S}^{'} = \rm{CME}(\widetilde{\textbf{Y}};\widetilde{\textbf{X}}).
\end{equation}

\subsection{Training Objectives of The Cross-modal Extractor}
To achieve coherent alignment between speech and text features and thereby facilitate a unified cross-modal representation that captures the essence of both modalities, we have formulated specialized loss functions at both the token and modal levels. In the subsequent sections, we will elaborate on these unique loss functions within the framework of our multi-task learning approach.

\subsubsection{Token-level loss} In the token-level training, we aim for the model to learn the context dependencies within text and speech sentences. Building upon previous work~\cite{wei2022conversational} and drawing inspiration from data2vec2.0~\cite{baevski2022efficient}, we no longer differentiate between text and speech intermediate features. For both speech and text modes, we consistently employ the method of predicting the masked portions of the features to enable the model to learn the context relationships within the sentence. \rev{In accordance with the method employed by previous work~\cite{yao2022tessp}, we up-sample the text. Specifically, we up-sample the characters using alignment information extracted from the ASR data.}

Concretely, we randomly mask 30\% of the speech features $\widetilde{\textbf{X}}$ and text features $\widetilde{\textbf{Y}}$ to obtain the masked features encoded by the cross-modal encoder $\widetilde{\textbf{X}}^{m} = \{x^m_1, x^m_2, ...,  x^m_T\}$ and $\widetilde{\textbf{Y}}^{m}=\{y^m_1, y^m_2, ...,  y^m_T\}$. Model training aims to predict the masked tokens using the remaining tokens. When predicting the masked text or speech, the features of the other mode will also be input into the model as a condition. Consequently, when predicting text sequences, our objective is to minimize the following negative logarithmic functions:
\begin{equation}
\begin{aligned}
  & \mathcal{L}_{speech} = - \sum_{t\in \mathcal{M}}(\text{log}{~p_{\theta_0}(x_t|x^m_t, \widetilde{\textbf{Y}}^{m})}),
    \label{eq:speech}
\end{aligned}
\end{equation}
\noindent where $\theta_0$ is trainable parameters in the model, $x_t$ is the target feature, and $\mathcal{L}_{speech}$ is the token-level loss of the speech. 

Similarly, the loss function of the speech encoder is defined as
\begin{equation}
\begin{aligned}
  & \mathcal{L}_{text} = - \sum_{t\in \mathcal{M}}(\text{log}{~p_{\theta_0}(y_t|y^m_t, \widetilde{\textbf{X}}^{m})}),
    \label{eq:speech2}
\end{aligned}
\end{equation}
\noindent where $y_t$ is the target feature, and $\mathcal{L}_{text}$ is the token-level loss of the speech. In line with the approach employed by data2vec, we utilize L1 Loss for both speech and text training.

\subsubsection{Modal-level loss}
In addition to the token-level loss, we also define a loss function at the modal level. We aim to learn the correlation between speech and its transcripts through the modal-level loss. Specifically, drawing inspiration from ~\cite{liu2021opt}, we randomly mask all tokens of the text or speech sentence with a certain probability 30\%, allowing the model to learn the corresponding representation through the input of another mode. When the text mode is masked, the input to the cross-modal feature encoder takes the following form:

\begin{equation}
\textbf{S} = \rm{CME}(\widetilde{\textbf{X}};\textbf{O}),
\end{equation}
or

\begin{equation}
\textbf{S}^{'} = \rm{CME}(\textbf{O};\widetilde{\textbf{Y}}),
\end{equation}
where $\textbf{O}$ represents the zero vector with the same length as the original vector.

Given that speech sequences are typically longer than text, we upsample the text features to equalize the feature lengths, thereby facilitating effective feature exchange between the two modalities. We incorporate an additional CTC loss~\cite{graves2012connectionist}, as utilized in~\cite{zhang2022speechlm} to enhance the inter-modal correspondence. This enables better time-series alignment between speech and text vectors. In our model, we utilize $\textbf{Y}$ as the CTC training target and employ text features $\widetilde{\textbf{Y}}$ and speech features $\widetilde{\textbf{X}}$ as the input. Incorporating CTC loss strengthens the alignment and bolsters the decoding performance in downstream ASR tasks.

\subsubsection{Total Extractor Loss}
We integrate the aforementioned token-level loss and modal-level loss functions to form the final loss function. Through multi-task learning, the cross-modal feature extractor can learn the context information within each mode and the mapping relationship between the two modes. The final loss function can be expressed as
\begin{equation}
    \mathcal{L}_{\rm{ext}} = {\alpha}\mathcal{L}_{\rm{CTC}}+{\beta}\mathcal{L}_{\rm{speech}}+{\gamma}\mathcal{L}_{\rm{text}}.
\end{equation}
In the final loss function, $\alpha$, $\beta$, and $\gamma$ are manually set parameters to control the weight of each loss component. Initially, a larger weight is assigned to $\alpha$ to expedite the mapping of speech and text into a common space. Subsequently, the weights of the three losses are balanced to enable the model to fully learn the inter-modal information. During the training of the cross-modal representation extractor, the parameters of the speech and text encoders are kept fixed. And when inference, we only use speech features as input and generate $\textbf{S}$ through a cross-modal encoder without feeding text input.

\section{The Conversational CVAE Module}
Inspired by~\cite{liang2021modeling}, we introduce a Conditional Variational Autoencoder (CVAE) module to extract conversation-related information from cross-modal vectors, further filtering out irrelevant information for conversational speech recognition and avoiding interference caused by lengthy historical information. Here, we feed the output of the cross-modal extractor $\textbf{S}_{role}$ and  $\textbf{S}_{topical}$, which is generated from $\widetilde{\textbf{X}}_{role}$ and $\widetilde{\textbf{X}}_{topical}$, into the CVAE module, and obtain a conversational representation containing role preference information and topical coherence information. 

The application of the CVAE method to obtain text representation has been extensively utilized across various fields~\cite{su2018variational,liang2021modeling}. By leveraging the VAE module and conditional information, the input features of the prenet are mapped to vectors containing information relevant to the target text, thus resulting in a more accurate representation of the target vector.

The CVAE module comprises a target text encoder and two Latent Variational Modules (LVMs). We employ the LVMs to extract conversational representations and feed them into the ASR model's decoder to capture topical and role context in conversation. When training, the cross-modal representation and the target text $\textbf{Y}_k$ will be fed into the prenet and postnet of LVMs, respectively. Then, we will use KL divergence to align these contextual cross-modal representations with the target text representation space. This implicit alignment enables the model to learn the relationships between contextual features and the text it aims to recognize. We will introduce the details of the CVAE model in this section.

\subsection{Input of LVMs}
When the input of LVM is a cross-modal feature, we feed $\widetilde{\textbf{X}}_{role}$ and $\widetilde{\textbf{X}}_{topical}$ into the cross-modal extractor to get the role representation $\textbf{S}_{role}$ and topical representation $\textbf{S}_{topical}$. In this configuration, the pre-trained language model receives no input. As a workaround, we generate a zero vector of equivalent length to the speech features and feed it, along with the speech features, into the cross-modal encoder:
\begin{align}
\textbf{S}_{role} &= \rm{CME}(\widetilde{\textbf{X}}_{role};\textbf{O}), \\
\textbf{S}_{topical} &= \rm{CME}(\widetilde{\textbf{X}}_{topical};\textbf{O}).
\end{align}
When the input is text feature, we send the contextual text $\textbf{Y}_{role}$, $\textbf{Y}_{topical}$ into the LVM text encoder, and get the $\textbf{S}^{text}_{role}$, $\textbf{S}^{text}_{topical}$. At the same time, the transcript of the current speech $\textbf{Y}_k$ is also an input of the postnet.

\subsection{Latent Variational Module}


As illustrated in Fig.~\ref{fig:frame}, each Latent Variational Module (LVM) comprises a prenet and a postnet. The role-specific LVM is designed to learn a role preference vector, denoted as $\textbf{V}_{role}$, while the topical LVM focuses on acquiring a topical coherence vector represented as $\textbf{V}_{topical}$. These vectors serve as latent variables, capturing context-specific nuances and topical coherence within the conversation.
In scenarios where the CVAE model processes text input, we introduce an additional Transformer block, termed the ``LVM text encoder.'' This block is responsible for extracting text features, which are subsequently provided to the LVM for learning the corresponding latent variables.

\subsubsection{LVM text encoder}
The text encoder in the LVM comprises multiple transformer layers, which are employed when the LVM model takes in text input. During this process, the text is first embedded into words and then transformed into a high-dimensional text feature by the text encoder. Importantly, all text inputs are processed through the same LVM text encoder to ensure consistency across the model:
\begin{align}
    \textbf{S}^{text}_{role} &= \rm{TextEnc}(\rm{Embedding}(\textbf{Y}_{role})), \\
    \textbf{S}^{text}_{topical} &= \rm{TextEnc}(\rm{Embedding}(\textbf{Y}_{topical})).
\end{align}
The variational representation $\textbf{a}_{topical}$ and $\textbf{a}_{role}$ are obtained by applying mean-pooling to the vectors generated by the word embedding operation (Embedding) and LVM text encoder (TextEnc) on the time dimension. This pooling process allows us to generate fixed-length vectors that capture the conversation's contextual information and topical coherence.

\subsubsection{Role LVM}
To generate the role preferences in the conversation, we utilize the variational representation $\textbf{a}_{role}$ obtained by mean-pooling the historical speech representations $\textbf{S}_{role}$ of the current speaker up to the $k$-th sentence, as represented by the role context information $\textbf{X}_{role}$ and $\textbf{Y}_{role}$. We model the role preferences using an isotropic Gaussian distribution, which has been shown to be effective in Wang~\etal~\cite{wang2019t}. We effectively capture the current speaker's role preferences by modeling the distribution based on historical role preferences and corresponding targets. These captured preferences are subsequently integrated into the latent variables within the role-specific LVM:
\begin{equation}
     p_{\theta}(\textbf{V}_\text{role}|\textbf{S}_\text{role})\sim N(\mu_\text{role}, \sigma^{2}_\text{role}\textbf{I}),
\end{equation}
\noindent where $\textbf{I}$ denotes the identity matrix, $\theta$ stands learnable parameters in prenet. Note that $\mu_\text{role}$ and $\sigma^{2}_\text{role}$ are calculated as follows:
\begin{align}
        \mu_\text{role} &= \rm{Lin}^{role}_{\theta}(\textbf{a}_{role}), \\
        \sigma_\text{role} &= \rm{Softplus}(Lin_{\theta}^{role}(\textbf{a}_{role})),
\end{align}
\noindent where the role preference vectors $\textbf{V}_{role}$ are obtained from a linear layer \texttt{Lin} and a \texttt{Softplus} activation function, which transforms the input into a high-dimensional latent space. Specifically, the prenet models the historical role preference characteristics in the conversation through a Gaussian distribution, while the postnet models the current role preference. To make the function of the prenet approach the postnet, we use KL divergence to measure the difference between the two distributions, as described in~\cite{sohn2015learning}.

The distribution function of the postnet is defined as follows:
\begin{equation}
     q_{\phi}(\textbf{V}^{p}_\text{role}|\textbf{S}_\text{role}, \textbf{Y}_{k})\sim N(\mu^{\prime}_\text{role}, \sigma^{\prime2}_\text{role}\textbf{I}),
\end{equation} 
and
\begin{align}
        \mu^{\prime}_\text{role} &= \rm{Lin}_{\phi}^{role}(\textbf{a}_{role}, \textbf{a}_{y}), \\
        \sigma^{\prime}_\text{role} &= \rm{Softplus}(\rm{Lin}_{\phi}^{role}(\textbf{a}_{role}, \textbf{a}_{y}))
\end{align}

\noindent Here, $\textbf{a}_{y}$ is the vector obtained by sending $\textbf{Y}_k$ into the LVM text encoder, $\phi$ is the learnable parameters in postnet.

The aforementioned processes are executed during the model's training phase. However, minimizing dependency on current recognition results during the decoding stage is crucial. We utilize the vector output from the preceding network layer to represent character preference features to accomplish this.

\subsection{Topical LVM}
We utilize a method similar to the role preference LVM to model topical consistency information in the conversation. Specifically, we use the topical coherence vector $\textbf{S}_{topical}$ to define an isotropic Gaussian distribution as follows:
\begin{equation}
     p_{\theta}(\textbf{V}_\text{topical}|\textbf{S}_\text{topical})\sim N(\mu_\text{topical}, \sigma^{2}_\text{topical}\textbf{I}),
\end{equation}
\noindent Here, $\textbf{I}$ denotes the identity matrix, $\theta$ stands learnable parameters in prenet. And 
\begin{align}
     \mu_\text{topical} &= \rm{Lin}^{topical}_{\theta}(\textbf{a}_{topical}), \\
     \sigma_\text{topical} &= \rm{Softplus}(Lin_{\theta}^{topical}(\textbf{a}_{topical})),
\end{align}
where $\textbf{a}_{topical}$ is obtained by mean-pooling the historical speech representations $\textbf{S}_{topical}$.

The prenet models the historical topical consistency characteristics in the conversation through Gaussian distribution, and the postnet models the current topical consistency:
\begin{equation}
     q_{\phi}(\textbf{V}_\text{topical}|\textbf{S}_\text{topical}, \textbf{Y}_{k})\sim N(\mu^{\prime}_\text{topical}, \sigma^{\prime2}_\text{topical}\textbf{I}),
\end{equation} 
and
\begin{align}
    \mu^{\prime}_\text{topical} &= \rm{Lin}_{\phi}^{topical}(\textbf{a}_{topical}, \textbf{a}_{y}), \\
    \sigma^{\prime}_\text{topical} &= \rm{Softplus}(\rm{Lin}_{\phi}^{topical}(\textbf{a}_{topical}, \textbf{a}_{y})).
\end{align}

The CVAE model will be trained together with the ASR system. During the decoding phase, only cross-modal representations are used as the input, eliminating the need for explicit transcript recognition. As demonstrated in our prior work~\cite{wei2022conversational}, the LVMs can be configured to accept either historical text embeddings or the cross-modal representations extracted by the aforementioned cross-modal extractor. 

\section{Experimental Setup}
\subsection{Dataset}
We evaluate our proposed method on two Chinese conversation datasets: MagicData-RAMC~\cite{yang2022open} and HKUST~\cite{liu2006hkust}. The HKUST dataset comprises telephone conversation recordings, while the MagicData-RAMC dataset consists of microphone conversation recordings captured in a quiet environment. To facilitate the effective extraction of role-based features, we re-segment the sentences according to speaker transitions, ensuring an alternating pattern between the two speakers. To enhance the diversity of the training data, we perform speed variation operations on the speech data from both datasets' training sets, specifically applying 0.9$\times$ and 1.1$\times$ speed changes. Detailed descriptions of the two employed datasets are as follows.

\subsubsection{MagicData-RAMC}
The MagicData-RAMC dataset~\cite{yang2022open} comprises 180 hours of Chinese conversational speech data, distributed as 150 hours for the training set, 20 hours for the development set, and 10 hours for the test set. The dataset features conversations from 663 speakers. Recordings were conducted in a quiet room, ensuring a noise level below 40dB during data collection. Speech data was captured using Android or Apple devices stored in a 16kHz, 16-bit format. The dataset encompasses 351 conversations, each centered around a specific topic.
The conversations encompass 15 distinct topics, such as the humanities, environment, family, sports, and more, thereby offering a comprehensive array of scenarios and subject matter.

\subsubsection{HKUST}
The HKUST dataset~\cite{liu2006hkust} comprises 200 hours of Mandarin Chinese conversational speech data, with a separate allocation of 60 minutes for the development set. It includes 1,206 conversations from 2,100 speakers, each lasting approximately 10 minutes. The development set consists of 12 conversations involving 24 speakers. Like the MagicData-RAMC dataset, the HKUST dataset covers a broad range of topics, with each conversation centering around a specific theme. All speech data is collected from phone calls and stored in an 8-bit, 8kHz format.  

\subsection{Implementation Details}
\subsubsection{Pre-trained models}
We employ the open-source Chinese HuBERT pre-trained base model\footnote{https://github.com/TencentGameMate/chinese\_speech\_pretrain} as the HuBERT speech encoder, adhering to the model configuration outlined in~\cite{hsu2021hubert}. The HuBERT base model comprises 12 transformer layers, each containing 768 nodes.

Furthermore, we train a data2vec model using the WenetSpeech train\_l dataset. This model is trained on the fairseq framework~\cite{ott2019fairseq}. Most of the model's configuration aligns with the base configuration in data2vec, comprising 12 transformer layers with 768 nodes each. However, to accommodate the data type of WenetSpeech, we modify certain parameters, such as reducing the minimum sentence length requirement and adjusting the number of warmup steps. These alterations enable the model to better adapt to the sentence length distribution and the larger scale of the new dataset.

\subsubsection{Cross-modal extractor}
During the extractor training process, we freeze the parameters of both the language and speech pre-trained models. The cross-modal encoder comprises three transformer layers. The text embedding vector obtains a high-dimensional text representation through the pre-trained language model during training. However, to reduce computational complexity during inference, we remove the pre-trained language model from the extractor and instead directly employ the zero vector combined with the input features of speech.

\subsubsection{CVAE based conversational ASR}
The CVAE-based conversational ASR architecture comprises a 12-layer Conformer encoder and a 6-layer transformer decoder. To incorporate historical information from the conversation, we utilize the enhanced decoder as described in Section II. The LVM text encoder comprises two layers of transformer blocks, as depicted in Fig.~\ref{fig:frame}. When the input of the LVM is a cross-modal representation, the LVM text encoder is removed.

\subsubsection{Features and tools}
We utilize raw wave files as the speech input for both the cross-modal extractor and the ASR model. The output from the speech encoder in the cross-modal extractor is fed into the cross-modal encoder and the ASR's transformer encoder. The cross-modal feature extractor and the ASR model are trained using the same supervised data, with the speech undergoing 0.9 and 1.1 ratio speed perturbations and 
SpeAugment~\cite{park2019specaugment}. To accommodate the input format of the pre-trained model, all speech data is uniformly converted to a 16 kHz sampling rate.

We pre-train the cross-modal extractor and utilize input without contextual information to initialize the ASR model. This approach prevents the model from overemphasizing historical information. Subsequently, we train the ASR model using the current speech-transcript pair, conversational role preferences, and topical coherence features. All models are trained using the open-source tool ESPnet~\cite{watanabe2018espnet}. Additionally, we employ the pre-training interface s3prl~\cite{yang2021superb} to convert the features of the speech pre-trained model.

\subsubsection{Baselines}
We employ a Conformer ASR model as the baseline, which comprises 12 layers with 512 nodes and 4 attention-head Conformer encoders, as well as 6 layers with 512 nodes and 4 attention-head Transformer decoders. Additionally, we use the data2vec pretrained Conformer ASR model~\cite{baevski2022data2vec} and the text-based CVAE model~\cite{wei2022conversational} from our previous work as supplementary baseline models for this study. The configurations of the CVAE model remain consistent with those in our previous work~\cite{wei2022conversational}.

\begin{table*}[htbp]
\centering

\caption{Comparison of CER (\%) for various models on two datasets. The \#Sentences column indicates the number of historical sentences utilized as input for the ASR model; ``0" implies that only the cross-modal representation of the current sentence is used. For CVAE models with CRM, the best performance is achieved using just one history sentence.}
{
\label{tab:total_results}
\begin{tabular}{cccccccccc}
\toprule
\multirow{2}{*}{\textbf{\#}} &
  \multirow{2}{*}{\textbf{Model}} &
  \multirow{2}{*}{\textbf{\#Sentences}} &
  \multirow{2}{*}{\textbf{Speech Pretrained Model}} &
  \multirow{2}{*}{\textbf{Modality of Pre/Postnet}} &
  \multirow{2}{*}{\textbf{HKUST/dev}} &
  \multirow{2}{*}{\textbf{RMAC/test}} \\ 
  \\ \hline
1 &
  Conformer-ASR~\cite{gulati2020conformer} &
   &
 - &
 - &
  20.3 &
  18.6 \\ 
2 &
  Pretrained-Conformer-ASR~\cite{baevski2022data2vec} &
   &
  data2vec &
  - &
  20.0 &
  16.0 \\ 
  3 &
  CVAE~\cite{wei2022conversational} &
  3 &
  - &
  text/text &
  19.3 &
  17.6 \\ 

4 &
  H-Transformer~\cite{masumura2021hierarchical} &
  3 &
  - &
  - &
  20.1 &
  18.3 \\ 

5 &
  Long-Context (reported)~\cite{hori2021advanced} &
   &
  - &
  - &
  17.3 &
  - \\ 

6 &
  Long-Context (reproduced) &
   &
  - &
  - &
  19.5 &
  15.8 \\ \hline

  7 &
  \multirow{3}{*}{CRM} &
  0 &
  data2vec &
  - &
  19.1 &
  15.1 \\ 
  
  8 &
  &
  1 &
  data2vec &
  - &
  18.7 &
  14.9 \\ 
  
  9 &
  &
  3 &
  data2vec &
  - &
  19.3 &
  16.2 \\ \hline
  
  10 &
  \multirow{3}{*}{CVAE} &
  3 &
  data2vec &
  text/text &
  19.0 &
  17.2 \\ 
  11 &
  &
  1 &
  data2vec &
  cross\_modal/text &
  19.6 &
  16.5 \\ 
  
  12 &
  &
  3 &
  data2vec &
  cross\_modal/text &
  18.7 &
  15.3 \\ \hline
  
  13 &
  \multirow{6}{*}{CVAE+CRM} &
  1 &
  data2vec &
  cross\_modal/text &
  19.4 &
  16.1 \\
  
  14 & 
  &
  3 &
  wav2vec2.0 &
  cross\_modal/text &
  19.3 &
  16.1 \\
  
  15 & 
  &
  3 &
  HuBERT &
  cross\_modal/text &
  18.9 &
  15.2 \\
  
  16 & 
  &
  3 &
  data2vec &
  cross\_modal/text &
  \textbf{18.5} &
  \textbf{14.3} \\

  17 & 
  &
  3 &
  data2vec &
  cross\_modal/cross\_modal &
  20.5 &
  18.4 \\

  18 & 
  &
  3 &
  data2vec &
  text/text &
  18.7 &
  16.3
  \\ \toprule
\end{tabular}}
\end{table*}

\section{Experimental Results}
We report the experimental results on two datasets and analyze the impact of different pre-trained models, various decoding methods, and additional language information. 
\subsection{Main Results}
Table~\ref{tab:total_results} presents the experimental results of our approach, which integrates cross-modal features and the CVAE conversational module. In the Model column, ``Conformer-ASR'' and ``Pretrained-Conformer-ASR'' are our baseline models, ``CRM'' indicates the use of the cross-modal extractor, and ``CVAE'' denotes the employment of conversational representations extracted using the LVM. In the CVAE model, the input for the prenet and postnet could be either cross-modal vector or text. 
The \#Sentences column specifies the number of historical sentences used to input the ASR model. In the CRM model, the cross-modal representation $\textbf{S}_{context}$ is directly fed into the decoder, while in the CVAE model, the conversational representations ($\textbf{V}_{role}, \textbf{V}_{topical}$) extracted by the CVAE module are fed into the decoder. In the CVAE+CRM model, the three features mentioned above are concatenated together and then fed into the decoder. We report the results for ASR models based on the FBank features (Model 1) and those based on the text conversation features from our previous work (Model 3). 
In addition, we also reproduce two models for comparison, including H-Transformer~\cite{masumura2021hierarchical} (Model 4) and Long-Context~\cite{hori2021advanced} (Model 6). Note that the reported result in~\cite{hori2021advanced} is also listed as Model 5.

Notably, Model 16, integrating both cross-modal and conversational features, demonstrates the lowest character error rate (CER) compared to the sentence-level ASR models and those reliant on text-based or cross-modal features alone.
Specifically, this model achieves an 8.8\% relative CER reduction compared to the Conformer-ASR baseline (Model 1). It also attains 3.1\% and 4.1\% error rate reductions compared to the text/text modality CVAE (Model 3) and CRM (Model 8) models, respectively. A similar phenomenon can be observed on the MagicData-RAMC dataset, which exhibits relative CER reductions of 23.1\%, 7.7\%, and 18.7\% compared to the aforementioned model categories. In comparison to the H-Transformer model (Model 6), which directly concatenates historical speech and text, the CVAE model reduces the CER by up to 3\%, demonstrating that the CVAE model can effectively map historical conversational context into more precise semantic representations. The HKUST dataset may be sensitive to hyperparameters such as input data order, learning rate, and training scale~\cite{kun2022Leveraging}. Consequently, our reproduction of Long-Context (Model 6) on the HKUST dataset achieved a character error rate of only 19.5\%. In contrast, our proposed method outperforms the reproduction of this method on the RMAC dataset, reducing CER from 15.8\% to 14.3\%. These results confirm that a speech recognition architecture enhanced with long-context conversational cues, cross-modal features, and conversational representations delivers superior performance.
The model of using conversational features to augment cross-modal representation addresses the possible error propagation from solely using textual features. It enables the system to leverage extended conversational context better. When the learning objective of CVAE is cross-modal representation, the CVAE module can not learn representations that are helpful to the ASR system. On the other hand, when the input and output of the CVAE module are both text, the decrease in CER is slightly less than that of the model with cross-modal representation as input.
\subsection{Influence of Conversation History Length on the Cross-modal Extractor}
Here, we investigate the effect of varying the length of historical conversation input for the cross-modal extractor. To achieve this, we concatenate the cross-modal representations of previous sentences with the representation of the current utterance. A comparative analysis of Models 7, 8, and 9 reveals that shorter spans of historical context consistently result in better recognition performance across all pre-trained models. This observation supports our previous hypothesis that an overload of historical data may dilute the model's focus on pertinent information, adversely affecting the recognition of the current sentence.


\subsection{Variability in Input Features for LVM Modules} 
We further extend our analysis to investigate the implications of different input features for the LVM. In previous work, textual representations exclusively served as inputs for both the postnet and prenet components of the LVM. In our current study, we diversify the input feature space by substituting one or more features with cross-modal representations. Fig.~\ref{fig:frame} outlines the implementation details: when the input to the LVM is textual, the text embeddings are processed through an LVM-specific text encoder to derive a context-rich text representation. In contrast, when employing cross-modal inputs, these inputs are fed directly into the LVM without further modification.


By comparing Models 10 and 12, which utilize cross-modal representations of historical speech to approximate the transcript of the current sentence, we observe that the model's recognition accuracy is significantly enhanced compared to methods using only text conversation features. This phenomenon can be attributed to the following reasons: on one hand, cross-modal representations contain both speech and text context information, allowing for better learning of the semantic relationship of text; on the other hand, it avoids error propagation caused by exclusively using text to represent conversational features. Concurrently, in Models 13-16, we incorporate the cross-modal representation of both the current and previous sentences into the decoder while adding conversational representations. We find that the recognition accuracy of ASR models is further improved by including cross-modal representations. This result suggests that the cross-modal representation of recent conversation may contain richer information, wherein the cross-modal information comprises more critical information than redundant information. Therefore, concatenating conversational representations can provide additional assistance for conversational speech recognition.

\subsection{Cross-modal Extractor with Various Pre-trained Speech Models}
In recent years, speech pre-training technology has made significant advancements. The features extracted by speech pre-trained models can replace traditional FBank and other features, thereby enhancing the recognition accuracy of speech recognition models. Furthermore, due to the robust feature extraction and representation capabilities of speech pre-trained models, we also utilize their outputs for cross-modal extractor training. 
We train cross-modal extractors based on three distinct pre-trained models and compare their final recognition error rates. All three models adopt the same configuration as the base model in fairseq, with consistent parameter values, and use a 10,000-hour WenetSpeech dataset~\cite{zhang2021wenetspeech} for pre-training. The models consist of 12 layers of transformer blocks, each with 768 nodes. During all fine-tuning processes, we freeze the parameters of the pre-trained models. Additionally, we compare the results of our method with the pre-trained model SpeechLM~\cite{zhang2022speechlm}, which also incorporates textual information into the pre-trained model. We fine-tune the SpeechLM model on the corresponding supervised datasets to ensure a fair comparison.

\begin{table}[htbp]
\centering
\caption{CER (\%) on RMAC test set of cross-modal representations with different speech pre-trained models. The input to the prenet in these configurations is cross-modal representation, while the postnet is fed with text embeddings.}
\label{tab:speech_model}
{
\begin{tabular}{c c c c}
\toprule
\textbf{Model}&\textbf{Pre-trained model} & \textbf{CER/RMAC} \\ \hline
\multirow{4}{*}{CVAE} & wav2vec2.0              & 16.8   \\ 
                     & HuBERT             & 15.7   \\ 
                     & data2vec            & \textbf{15.3}   \\
                     & SpeechLM              & 16.2   \\ \toprule
\end{tabular}
}
\end{table}

In ASR tasks, the performance of the three pre-training models aligns with the findings from other studies: the HuBERT model outperforms wav2vec2.0~\cite{yang2021superb}, and the data2vec model surpasses the HuBERT model~\cite{baevski2022data2vec}. From the cross-modal extractor experiments (Models 14, 15, and 16 in Table~\ref{tab:total_results}), we can draw a similar conclusion: HuBERT exhibits stronger capabilities than wav2vec2.0 in extracting semantic information, while data2vec's semantic extraction ability is superior to the other two models. The results in Table~\ref{tab:speech_model} support the same conclusion. This superiority might be attributed to data2vec's closer resemblance to text during speech pre-training and the lack of a need to map codebooks. Furthermore, by comparing the results of HuBERT and SpeechLM in Table~\ref{tab:speech_model}, we can conclude that our cross-modal extractor demonstrates excellent ability in extracting conversation-related cross-modal representations.


\subsection{Impact of Conversational Representation Length}
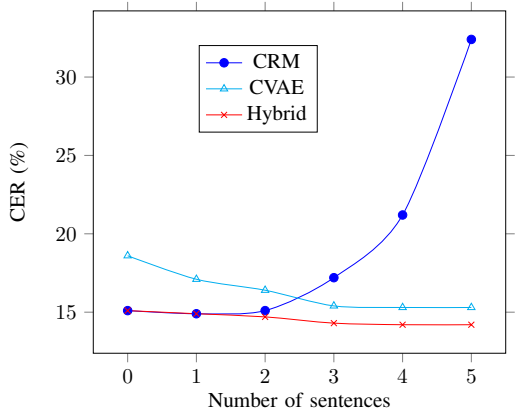
\begin{figure}[htbp]
\centering
\scalebox{0.8}{
\begin{tikzpicture}

\begin{axis}[
    xlabel=Number of sentences, 
    ylabel=CER (\%), 
    tick align=outside, 
    legend style={at={(0.4,0.9)},anchor=north} 
    ]

\addplot[smooth,mark=*,blue] plot coordinates {
    (0,15.1)
    (1,14.9)
    (2,15.1)
    (3,17.2)
    (4,21.2)
    (5,32.4)
};

\addlegendentry{CRM}

\addplot[smooth,mark=triangle,cyan] plot coordinates {
    (0,18.6)
    (1,17.1)
    (2,16.4)
    (3,15.4)
    (4,15.3)
    (5,15.3)
};
\addlegendentry{CVAE}

\addplot[smooth,mark=x,red] plot coordinates {
    (0,15.1)
    (1,14.9)
    (2,14.7)
    (3,14.3)
    (4,14.2)
    (5,14.2)
};
\addlegendentry{Hybrid}

\end{axis}
\end{tikzpicture}
}
\caption{CER vs. conversation history (number of sentences).}
\label{length}
\end{figure}

We posit that leveraging cross-modal conversational representations can enable more effective utilization of extended conversation history without sacrificing model performance. To empirically validate this hypothesis, we analyze the variations in CER as the length of conversation history input is extended. Evaluations are performed on the MagicData-RAMC dataset, and the findings are visualized in Fig. \ref{length}. Three distinct configurations are considered:
\begin{itemize}
    \item CRM: Only cross-modal representations are fed into the Automatic Speech Recognition (ASR) model.
    \item CVAE: In this case, only conversational representations are used as input to the ASR model.
    \item Hybrid: Both cross-modal and conversational representations are utilized as inputs to the ASR model.
\end{itemize}



As depicted in Fig.~\ref{length}, we find an intriguing trend in CER concerning the length of historical input. Models incorporating direct cross-modal representations initially exhibit a decline in CER, which subsequently deteriorates as the history length exceeds five sentences. This phenomenon suggests a significant degradation in the model's recognition capabilities under such conditions.

To mitigate this, we experiment with selectively feeding the decoder of the speech recognition model with only the cross-modal and conversational representations of the immediate previous sentence. By restricting the length of cross-modal representation to encompass just the current and previous sentences, we observe a noticeable enhancement in model performance. Specifically, on the MagicData-RMAC dataset, this configuration results in approximately a 6\% decrease in CER, aligning well with the trends observed for models relying solely on conversational representations.

From the above results, we can confirm our conjecture that using only cross-modal representations can interfere with the recognition of the current sentence's speech due to excessive historical information, while using conversational representations can avoid this phenomenon. Moreover, our analysis reveals that conversational and cross-modal representations are complementary rather than redundant. The incorporation of additional cross-modal features can indeed enhance the recognition accuracy of conversational ASR systems. This validates the merit of adopting a hybrid approach that synergistically combines both feature types, offering a more robust solution for handling long-context conversational data in ASR systems.

\subsection{Additional Language Information}
Table~\ref{LM} presents the experimental results comparing the effects of language models in three systems: 
A system reliant solely on data2vec features, a second leveraging only cross-modal features, and a third integrating both cross-modal and conversational features. To refine our approach, we restrict the cross-modal features to the current and immediate previous sentence in the conversation, while the conversational features are derived from the first three sentences. The systems utilizing cross-modal pre-trained models employ data2vec as their backbone. The input to the prenet in these configurations is cross-modal representation, while the postnet is fed with text embeddings.

\begin{table}[htbp]
\centering
\caption{The CER (\%) of using language models in different models on RMAC test set.}
{
\begin{tabular}{ccccc}
\toprule
\multicolumn{1}{c}{\textbf{\#}} &
  \multicolumn{1}{c}{\textbf{Model}} &
  \multicolumn{1}{c}{\textbf{LM}} &
  \multicolumn{1}{c}{\textbf{CER}} \\ \hline
1  & Data2vec\_Conformer (Pretrained) & - & 16.0          \\ 
2  & Data2vec\_Conformer (Pretrained) & Transformer LM & 15.7          \\ 
3  & CRM & - & 14.9          \\ 
4  & CRM & Transformer LM & 14.9          \\ 
5  & CVAE+CRM & - & \textbf{14.3} \\ 
6  & CVAE+CRM & Transformer LM & 14.4          \\ \toprule
\end{tabular}
}
\label{LM}
\end{table}

The results presented in Table~\ref{LM} suggest a nuanced relationship between language models and ASR performance. Specifically, when no conditional information is utilized (\#1 and \#2), language models provide a noticeable enhancement to the system's speech recognition capability. However, this advantage diminishes when cross-modal representations are incorporated, with the absolute change in recognition performance being a mere 0.1 (\#3 and \#4). Even more strikingly, the utility of language models is nearly nullified when both conversational and cross-modal features are leveraged (\#5 and \#6).

These results demonstrate the richness of the semantic information captured by our cross-modal and conversational representations. Notably, when employing a fusion of both feature types, our ASR model can extract semantic insights, thereby bolstering its speech recognition efficacy.


Concurrently, we observe that our conversational speech recognition model (\#5) incorporates additional LVM modules and a cross-modal extractor module compared to traditional speech recognition models. This is nearly equivalent to the parameter amount of the pre-trained model combined with the language model (\#2). However, our conversational speech recognition model achieves significantly improved recognition performance while using nearly the same number of parameters as the pre-trained model plus the language model. This underscores the effectiveness of our approach in optimizing conversational ASR systems.

\subsection{Comparison of Two Decoding Methods}
In Table~\ref{tab:fusion}, we present the CER results for the two different conditional information fusion strategies on the MagicData-RMAC dataset. As mentioned earlier, the speech pre-trained models employ HuBERT, with the input of the prenet being cross-modal and the input of the postnet being text embedding.

\begin{table}[htbp]
\centering
\caption{CER (\%) of different fusion strategies in different methods.}
\label{tab:fusion}
{
\begin{tabular}{c c c c}
\toprule
\textbf{Model}&\textbf{\#Sentences} & \textbf{Attention} & \textbf{Linear} \\ \hline
CRM      & 1         & \textbf{14.8}      & 14.9   \\ 
CRM     & 3          & 21.7     & \textbf{16.2}   \\ 
CVAE               & 3         & 15.7      & \textbf{15.3}   \\ \toprule
\end{tabular}
}
\end{table}

Our experiments demonstrate that the effectiveness of using attention layers as fusion strategies tends to deteriorate as sentence length increases. With only one sentence of cross-modal historical information, attention fusion performs slightly better than linear fusion. However, when using three sentences of cross-modal historical information, attention fusion performs significantly worse than linear fusion. A similar pattern is observed in experiments based on conversational representations.

This phenomenon occurs because excessively long historical information may interfere with the recognition of current speech, and the additional attention layer might allow the decoder to obtain more irrelevant information, exacerbating the distraction when inputting extended historical information. Our analysis of the attention distribution of the decoder for the same sentence with varying lengths of historical input revealed that as historical information lengthens, the attention weighting for the current sentence weakens considerably.

In experiments based on conversational representations, we reach the same conclusion. When the historical input of conversational representations comprises three sentences, linear fusion achieves higher recognition accuracy. This suggests that while attention fusion may have more parameters and a greater likelihood of capturing key information in history sentences, an overly strong attention mechanism might not be fully suitable for the fusion of conditional information. Alternatively, an additional attention layer might require further experiments to adjust the decoder's training objectives.

In future work, we will explore more suitable decoder attention fusion strategies and continue to optimize the conversational ASR system for improved performance.


\subsection{Ablation Study of Role and Topical Context Information}
We further investigate the influence of role and topical context information on the recognition results in Table~\ref{tab:role}. We observe that when only role or topical representation is employed, the final recognition result experiences a noticeable decline. In instances where the number of historical sentences is 3, topical features' impact surpasses role features. We attribute this to the role features utilizing context that is too distant ($\textbf{X}_{k-6}$). Although the role representation incorporates the speaker's information, it simultaneously weakens the connection with the current sentence~\cite{xiong2018session}. When both representations are combined, the model's CER is further reduced.

We also evaluate the Perplexity (PPL) of the language model in Table~\ref{tab:ppl}, incorporating both role information and topical information on the HKUST and RMAC datasets. When the training data $\textbf{Y}_k$ of the language model is supplemented with $\textbf{Y}_{role}$ and $\textbf{Y}_{topical}$, the reduction in PPL is comparable to the performance improvement observed in the ASR model.
\begin{table}[htbp]
\centering
\caption{CER (\%) of different CVAE information on RMAC test set.}
\label{tab:role}
{
\begin{tabular}{c c c}
\toprule
\textbf{Model} &\textbf{Context information} & \textbf{Results}  \\ \hline
\multirow{3}{*}{CVAE}      & role         & 16.2    \\ 
     & topical          & 15.6        \\ 
      & role\&topical         & \textbf{15.3}     \\  \hline
\multirow{3}{*}{CVAE+CRM}      & role         & 14.8    \\ 
     & topical          & 14.5        \\ 
      & role\&topical         & \textbf{14.3}   
      \\ \toprule
\end{tabular}
}
\end{table}

\begin{table}[htbp]
\centering
\caption{PPL of different role and topical information on HKUST and RMAC test set.}
\label{tab:ppl}
{
\begin{tabular}{c c c c}
\toprule
\textbf{Model} &\textbf{Context information} & \textbf{HKUST}  &\textbf{RMAC}\\ \hline
\multirow{4}{*}{Transformer LM}      & -         & 44.58 & 39.26    \\ 
      & role         & 41.37 &  36.81    \\ 
      & topical          & 38.85 &  33.22         \\ 
     & role\&topical         & \textbf{36.61}  & \textbf{29.72}      \\ \toprule
\end{tabular}
}
\vspace{-0.3cm}
\end{table}

\subsection{Comparison of Parameter and Real-time Factor}
For a fair comparison, we calculate the parameter quantities of different models and the real-time decoding factor on the RMAC test set in Table~\ref{LM2}.

\begin{table}[htbp]
\centering
\caption{Parameter number, real-time factor (RTF) and CER (\%) for different models in RMAC test set. The input to the prenet in these configurations is cross-modal representation, while the postnet is fed with text embeddings.}
{
\begin{tabular}{ccccc}
\toprule
\multicolumn{1}{c}{\textbf{Model}} &
  \multicolumn{1}{c}{\textbf{Parameter (M)}} &
  \multicolumn{1}{c}{\textbf{RTF}} &
  \multicolumn{1}{c}{\textbf{CER}} \\ \hline
Data2vec\_Conformer & 207.3 & 0.94 & 16.0          \\ 
Data2vec\_Conformer+LM & 258.6 & 1.24 & 15.5         \\ 
CRM & 216.9 & 1.21 & 14.9  \\ 
CRM+LM & 268.2 & 1.52 &14.9 \\
CVAE & 240.5 & 1.28 & 15.3  \\ 
CVAE+CRM & 240.5 & 1.32 & \textbf{14.3} \\ 
\toprule
\end{tabular}
}
\vspace{-0.2cm}
\label{LM2}
\end{table}

As the parameters of the cross-modal extractor are frozen during training, we can reuse the cross-modal representation to reduce the RTF. The increase in the number of parameters for our proposed method is relatively insignificant, and it even possesses nearly 20 million fewer parameters than the baseline model with the added language model (LM). Despite this, our approach demonstrates a substantial improvement in recognition accuracy.

By reusing historical representations, the RTF of our system is marginally slower than the baseline system. However, it remains essentially consistent with previous methods and does not significantly impact decoding efficiency.

\section{Conclusion}
This paper presents an innovative conversational ASR architecture that effectively recognizes speech within a conversational context using a CVAE module and cross-modal representation learning. We incorporate local and long contexts in conversational speech recognition without explicit error propagation and attention dilution. The proposed framework attains significant performance improvements on two challenging datasets, HKUST and MagicData-RAMC, showcasing its potential to enhance conversational speech recognition. By addressing the limitations of existing ASR systems in capturing conversational context, our work lays the foundation for future research and development in this area, aiming to develop more efficient, accurate, and context-aware ASR systems.










\bibliographystyle{IEEEtran}

\bibliography{mybib}

\newpage

\end{document}